\documentclass{PoS}
\usepackage{verbatim}
\usepackage{epsfig}

\title{The QCD equation of state and transition at finite temperature}

\ShortTitle{The QCD equation of state and transition at finite temperature}

\author{\speaker{Michael Cheng} for the HotQCD Collaboration\thanks{A. Bazavov, T. Bhattacharya, M. Cheng, N. H. Christ, C. DeTar, S. Ejiri, Steven Gottlieb, R. Gupta, U. M. Heller, K. Huebner, C. Jung, F. Karsch, E. Laermann, L. Levkova, C. Miao, R. D. Mawhinney, P. Petreczky, C. Schmidt, R. A. Soltz, W. Soeldner, R. Sugar, D. Toussaint, P. Vranas}\\
Physics Division, Lawrence Livermore National Laboratory\\
Email: \email{cheng24@llnl.gov}
        }
	
\abstract{We present the latest results for the equation of state and the
crossover transition in 2+1 flavor QCD
from the HotQCD Collaboration.  Bulk thermodynamic quantities - energy
density, pressure, entropy density, and the speed of sound - are calculated
on lattices with temporal extent $N_t = 8$ in the temperature range
$140~\textrm{MeV} < T < 540~\textrm{MeV}$.  We utilize two improved staggered
fermion actions, asqtad and p4, with the mass for the two degenerate light
quarks chosen to be $m_{ud} = 0.1 m_s$, corresponding to $m_\pi \approx 220~MeV$
for the lightest pion.  We also calculate observables that are sensitive to the
chiral and deconfing transitions - the light and strange quark number susceptibilities, the chiral condensate, and the renormalized Polyakov loop - finding 
that deconfinement and chiral symmetry restoration occur in the same narrow
temperature interval.
}

\FullConference{The XXVII International Symposium on Lattice Field Theory - LAT2009\\
		 July 26-31 2009\\
		 Peking University, Beijing, China}

\begin{document}

\section{Introduction}

Among the most important and fundamental problems in finite-temperature QCD are the
calculation of the bulk properties of hot QCD matter and the characterization of the nature
of the QCD phase transition.  Understanding finite temperature QCD also has direct application in
interpreting the results from heavy ion collision experiments.
Here, we present a lattice calculation of the QCD equation of state,
\textit{i.e.}, the pressure, energy density, entropy density, and speed of sound, at finite
temperature and vanishing chemical potential.  We also calculate quantities such as the 
chiral condensate, renormalized
Polyakov loop, and the light and strange quark number susceptibilities, which are related
to the chiral and deconfining aspects of the QCD transition.  For a more detailed discussion
of our results, see \cite{Bazavov:2009zn}.

\section{Simulation Details}

This calculation presents results with two different improved staggered fermion actions,
p4 and asqtad, with lattices with temporal extent $N_t = 8.$  Both the p4 and asqtad actions
eliminate the $\mathcal{O}(a^2)$ errors in bulk thermodynamic observables at high temperature, so there
are only small deviations from the asymptotic, ideal gas limit even at $N_t = 6$ and 8 \cite{Heller:1999xz}.
Symanzik-improved gauge actions are used with both the p4 and asqtad fermion actions.  For
the p4 action, a tree-level improved gauge action
is employed.  For the asqtad action, a one-loop  improved gauge action is utilized, with the
addition of tadpole improvement in both the gauge and fermion parts of the action.  Furthermore,
to reduce the effects of taste-symmetry breaking, fat-link smearing is implemented.  
The p4 action adds the three-loop staple into the fat link, while the asqtad action
adds terms up to the seven-link staple which minimize the taste-mixing terms in the fermion
action.

The finite temperature results presented here at $N_t = 8$ are obtained from lattices of size
 $32^3 \times 8$ generated using the RHMC algorithm.  These results are presented with 
 previously obtained results at $N_t = 6$ \cite{Cheng:2007tp,Bernard:2006nj}.  In the case of the asqtad action, some new measurements were also
made on $32^3 \times 6$ lattices.  In
addition, "zero temperature" calculations have also been performed on $32^4$ lattices where
the gauge coupling $\beta = 6/g^2$ and bare quark masses are chosen to be the same as
a corresponding ensemble at finite temperature.

These calculations were performed with two degenerate light quarks, and a heavier strange 
quark.  The strange quark mass is chosen to be close to its physical value, with the light quark 
mass of $m_{ud} = 0.1 m_s$.  The bare masses are fixed so that they lie on
a line of constant physics, \textit{i.e.}, the hadron masses remain fixed in physical units.  In our
case, $m_K$, $m_{\bar{s}s}$, and $m_\pi$ are used to fix the line of constant physics. This corresponds to a lightest pion of $m_\pi \approx 220$ MeV.
The line of constant physics differs slightly between the p4 and asqtad actions, with the 
strange pseudoscalar mass, $m_{\bar{s}s}$ approximately 15\% larger in the asqtad case
than with the p4 action.
In order to set the lattice scale in physical units, we use the quantities $r_0$ and $r_1$, which
are related to the shape of the heavy quark potential:
\begin{equation}
\left(r^2 \frac{dV_{\bar{q}q}(r)}{dr}\right)_{r=r_0} = 1.65;~~\left(r^2 \frac{dV_{\bar{q}q}(r)}{dr}\right)_{r=r_1} = 1.0.
\end{equation}
We use $r_0 = 0.469(7)$ fm. in order to translate our lattice data into physical units \cite{Gray:2005ur}.

\section{QCD Equation of State}
In determining bulk thermodynamic observables on the lattice, the basic quantity that we
calculate is the trace anomaly $\Theta^{\mu \mu}(T)$,
\begin{equation}
\frac{\Theta^{\mu \mu}(T)}{T^4} = \frac{\epsilon - 3p}{T^4} = T \frac{\partial}{\partial T} \left(\frac{p}{T^4}\right).
\end{equation}
\begin{figure}[bt]
\vspace{-0.3cm}
\includegraphics[width=0.47\textwidth]{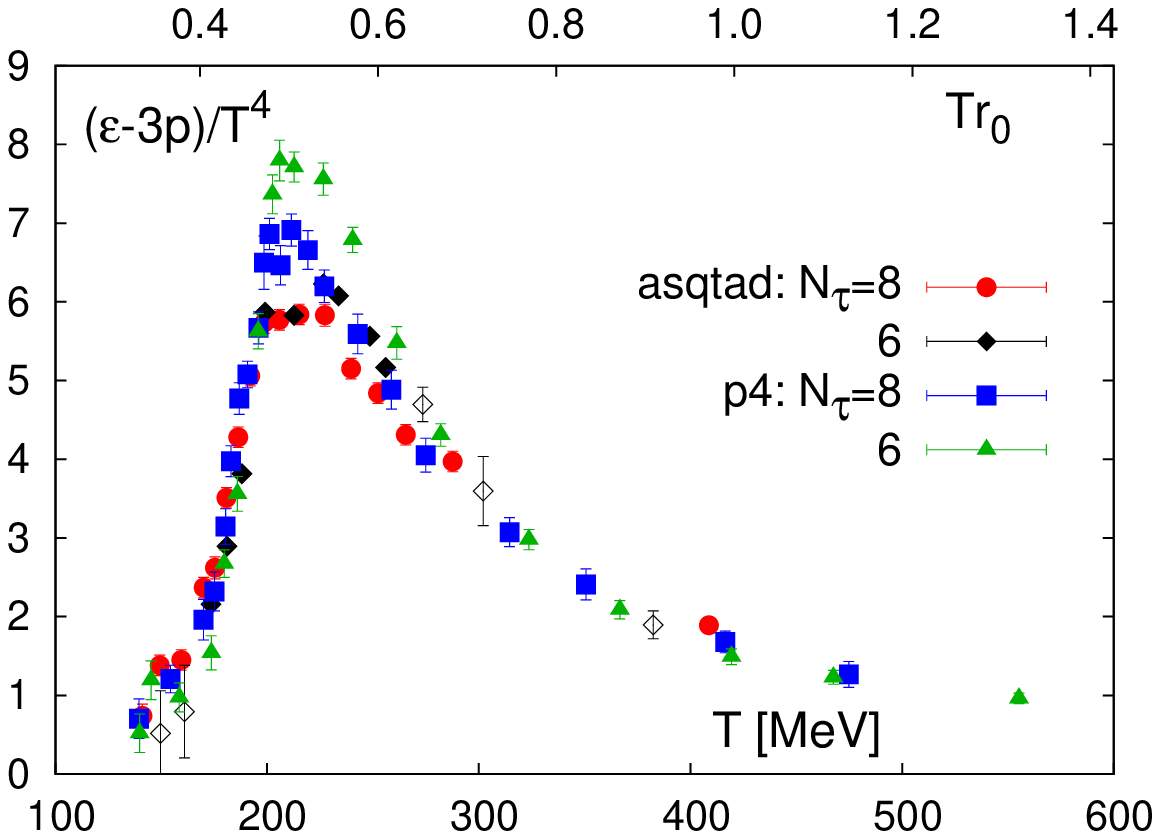}
\includegraphics[width=0.47\textwidth]{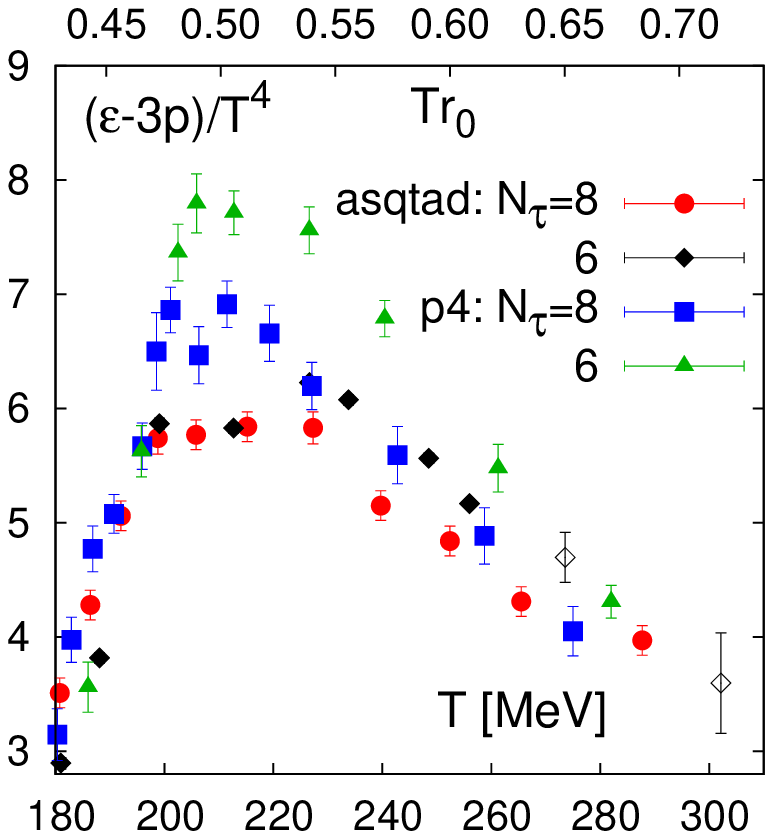}
\caption{On the left, the trace anomaly, \textit{i.e.}, $(\epsilon - 3p)/T^4$ for the p4 and asqtad actions at $N_t = 6$ and 8.  On the right, the same quantity, with the peak of the trace anomaly
shown in more detail.}
\label{fig:e-3p_all_center}
\end{figure}
\begin{figure}[b]
\includegraphics[width=0.47\textwidth]{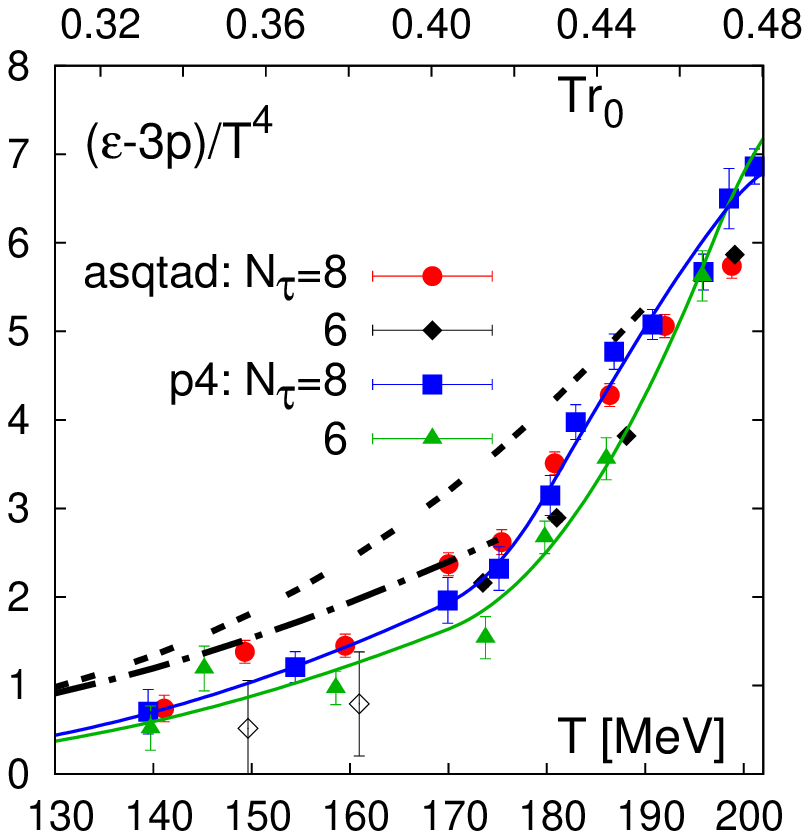}
\includegraphics[width=0.47\textwidth]{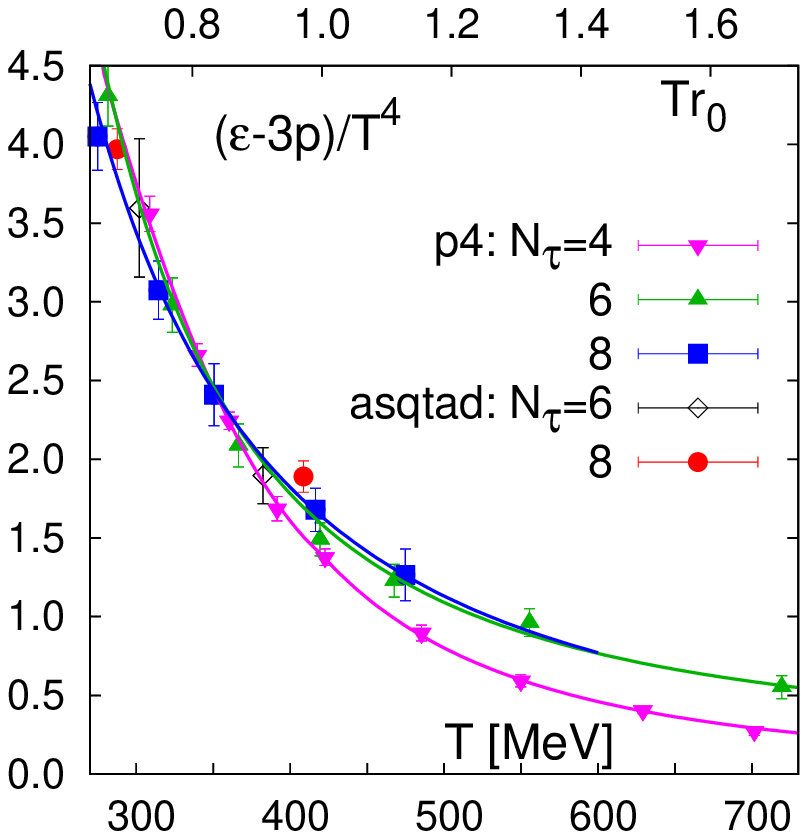}
\caption{On the left, the trace anomaly for the p4 and asqtad actions shown at low temperature,
with a comparison to the hadron resonance gas.  The dashed-dotted/dashed curve is the hadron resonance gas with a maximum resonance mass of $m_\textrm{max} = 1.5/2.5$ GeV.  On the right, the trace anomaly at high temperature.}
\label{fig:e-3p_low_high}
\end{figure}
The trace anomaly can be separated into contributions from two pieces.  One contribution comes
from the strange and light quark condensates, and vanishes in the chiral limit, \textit{i.e.},
$\hat{m}_l, \hat{m}_s \rightarrow 0,$
\begin{equation}
\frac{\Theta^{\mu \mu}_F(T)}{T^4} = -R_\beta R_m N_t^4 \left(2 \hat{m}_l \Delta \left<\bar{\psi} \psi\right>_l + \hat{m}_s \Delta\left< \bar{\psi} \psi\right>_s\right),
\end{equation}
and a piece that is non-vanishing in the chiral limit:
\begin{equation}
\frac{\Theta^{\mu \mu}_G(T)}{T^4} =  R_\beta N_t^4 \left( \Delta\left< s_G \right> - R_u \Delta \left<\frac{d (s_G + s_F)}{d u_0} \right>\right),
\end{equation}
where $R_\beta, R_m,$ and $R_u$ are nonperturbative beta functions which describe how the bare parameters change with scale:
\begin{equation}
R_\beta = T\frac{d \beta}{dT} = - a \frac{d \beta}{da};~~ R_m = \frac{1}{\hat{m}_l(\beta)} \frac{d \hat{m}_l(\beta)}{d \beta};~~ R_u = \beta \frac{d u_0(\beta)}{d \beta},
\end{equation}
and the notation $\Delta \left<X\right> = \left<X\right>_0 - \left<X\right>_T$ indicates a zero
temperature subtraction has been made.

Figure \ref{fig:e-3p_all_center} shows the results for the trace anomaly at
$N_t = 6$ and 8.  As one can see, there is quite good agreement between the two different actions,
particularly in the high temperature region ($T > 300$ MeV).  Figure \ref{fig:e-3p_all_center} also 
shows the peak region, where the differences between the two actions are largest.  In particular,
the peak for the p4 action is about 15\% higher than for the asqtad action.

Figure \ref{fig:e-3p_low_high} shows the trace anomaly at both low and high temperature.  Although the p4 and asqtad results are generally consistent at low temperature, 
we find that they (not surprisingly) do not agree with those of the hadron resonance gas model, as 
the hadron spectrum is distorted by heavier-than-physical quark masses
and taste symmetry violations.  At high temperature, we see that we have very good agreement
between the p4 and asqtad actions, as well as evidence for small lattice artifacts.
\begin{figure}[bt]
\vspace{-0.3cm}
\includegraphics[width=0.47\textwidth]{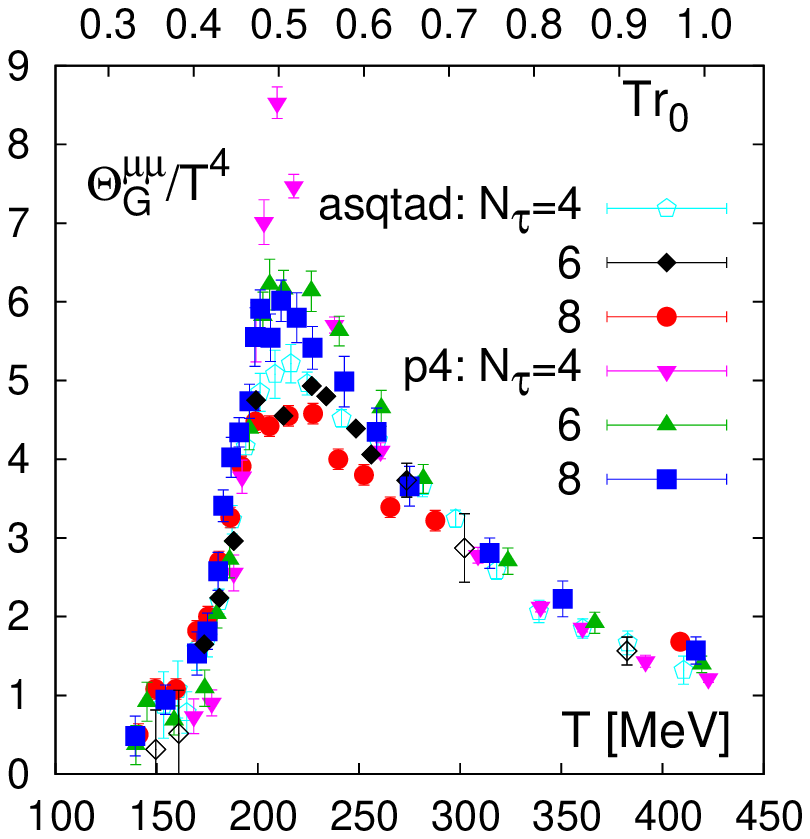}
\includegraphics[width=0.47\textwidth]{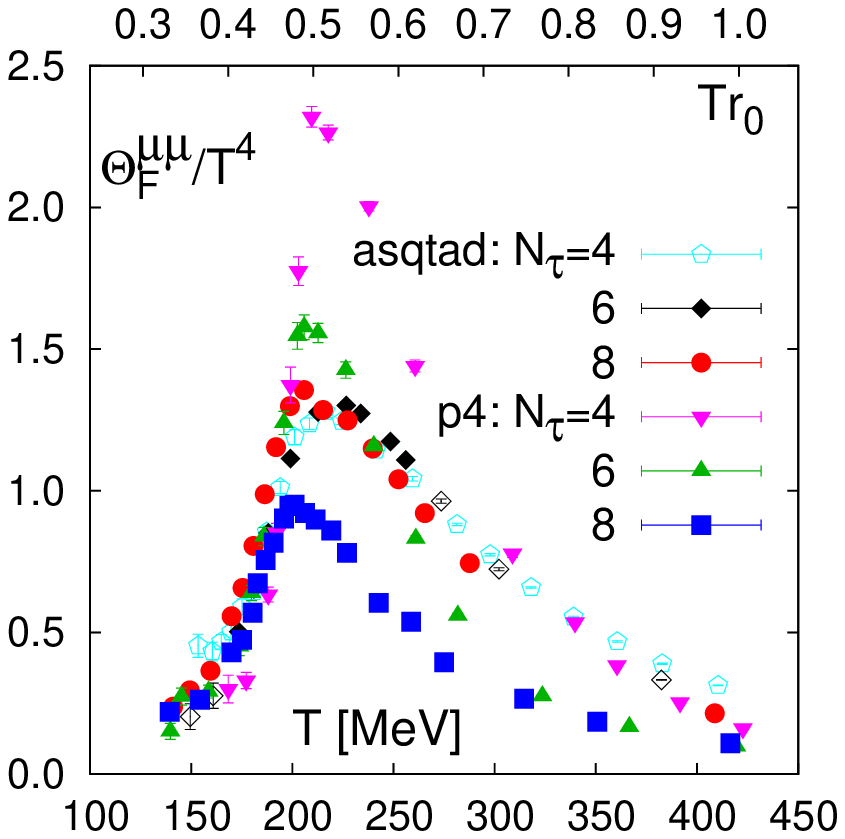}
\caption{Contributions to the trace anomaly from gluonic observables, on the left, and for
fermionic observables on the right, as defined in the text.  Results for both 
actions at $N_t = 4,6$, and $8$ are shown.}
\label{fig:Theta}
\end{figure}
Figure \ref{fig:Theta} shows the contributions to the trace anomaly from $\Theta^{\mu \mu}_G$
and $\Theta^{\mu \mu}_F$.  Although identifying these two terms as "gluonic" and "fermionic"
contributions is not accurate (as the dynamical quarks mix these two terms), one can
see that the gluonic observables dominate the trace anomaly.  In addition, whereas there is
little evidence for cut-off dependence in $\Theta^{\mu \mu}_F$ for the asqtad action, there is
significant cutoff dependence for p4.  This cutoff dependence can be traced to the
nonperturbative beta function $R_m$ at strong coupling.  Indeed, one can see that the cutoff
effects in $\Theta^{\mu \mu}_F$ are diminished at higher temperature.

From the trace anomaly, one can reconstruct the pressure via integration:
\begin{equation}
\frac{p(T)}{T^4} - \frac{p(T_0)}{T_0^4} = \int_{T_0}^T dT' \frac{1}{T'^5}\Theta^{\mu \mu}(T'),
\end{equation}
where $T_0$ is chosen to be a temperature sufficiently deep in the confined regime so that
$p(T_0)$ is small.  From the pressure and trace anomaly, it is straightforward to construct 
the energy density, entropy density, and speed of sound.

\begin{figure}[bt]
\vspace{-0.3cm}
\includegraphics[width=0.47\textwidth]{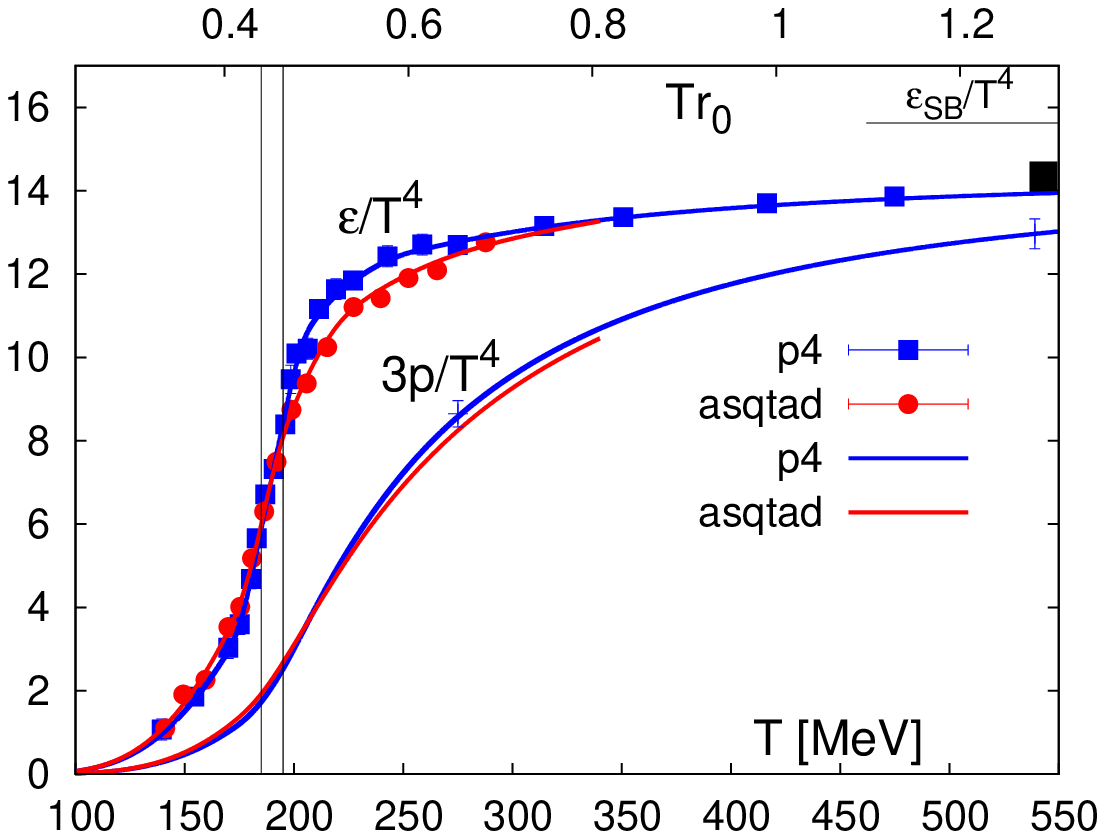}
\includegraphics[width=0.47\textwidth]{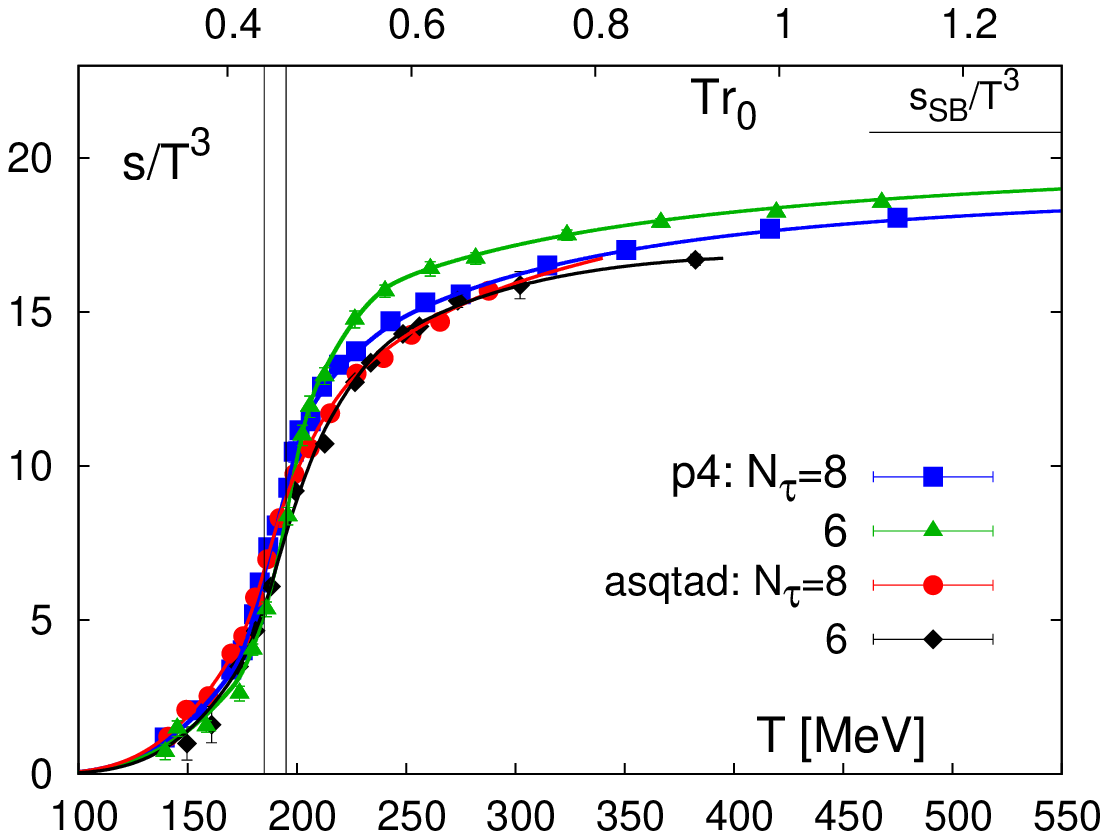}
\caption{On the left, a comparison of the energy density and three times the pressure for the
p4 and asqtad actions, at $N_t = 8$.  The black bar in the upper right hand corner indicates the
systematic offset from adding the hadron resonance gas value of the pressure at $T_0 = 100$ MeV.  On the right, the entropy density for both actions at $N_t = 6$ and 8. The crossover region of 185 MeV < T < 195 MeV is also shown on both plots.}
\label{fig:energy_pressure_entropy}
\end{figure}

Figure \ref{fig:energy_pressure_entropy} shows the energy density, three times the pressure, and
entropy density.  Our data show that the relative difference between the two actions is no more
than about 15\% in the low temperature region for $T > 150$ MeV, and falls to about 5\% for
$T > 200$ MeV.  For the p4 action, the cutoff effects between $N_t = 6$ and 8 are of similar size, while
no statistically significant cutoff dependence is seen for the asqtad action.

\begin{figure}[hbt]
\begin{minipage}{0.50\textwidth}
\includegraphics[width=\textwidth]{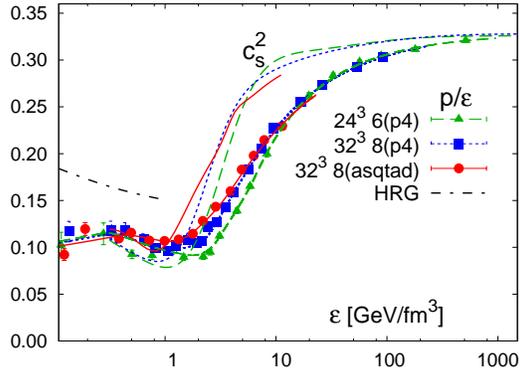}
\end{minipage}
\begin{minipage}{0.50\textwidth}
\caption{Pressure divided by the energy density ($p/\epsilon$) and the speed of sound squared 
($c_s^2$) .  Data points show the ratio of pressure to energy density, while colored dashed
curves show $c_s^2$ obtained from the interpolations of $\epsilon/T^4$ and $p/T^4.$  The
dashed-dotted line at low temperature shows the result for the hadron resonance gas model
with $m_{max} = 2.5$ GeV.} 
\label{fig:cs}
\end{minipage}
\end{figure}

Figure \ref{fig:cs} shows the ratio of the pressure and energy density, as well as the speed of 
sound.  Although there is qualitative agreement between the two actions, the speed of sound
significantly undershoots the hadron resonance gas value at low temperature.  The speed of
sound is given by:
\begin{equation}
c_s^2 = \frac{dp}{d \epsilon} = \epsilon \frac{d(p/\epsilon)}{d \epsilon} + \frac{p}{\epsilon}.
\end{equation}
\vspace{-0.7cm}
\vspace{-1cm}
\section{QCD Transition}
\begin{figure}[t]
\includegraphics[width=0.47\textwidth]{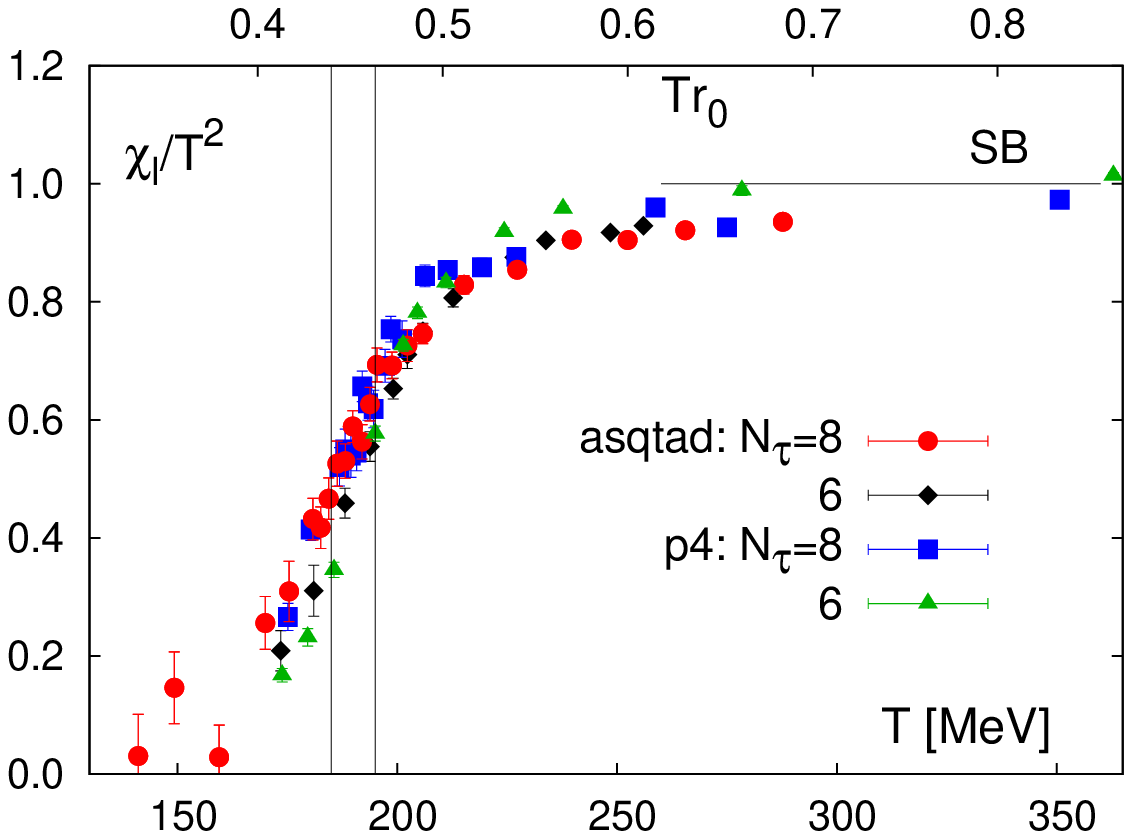}
\includegraphics[width=0.47\textwidth]{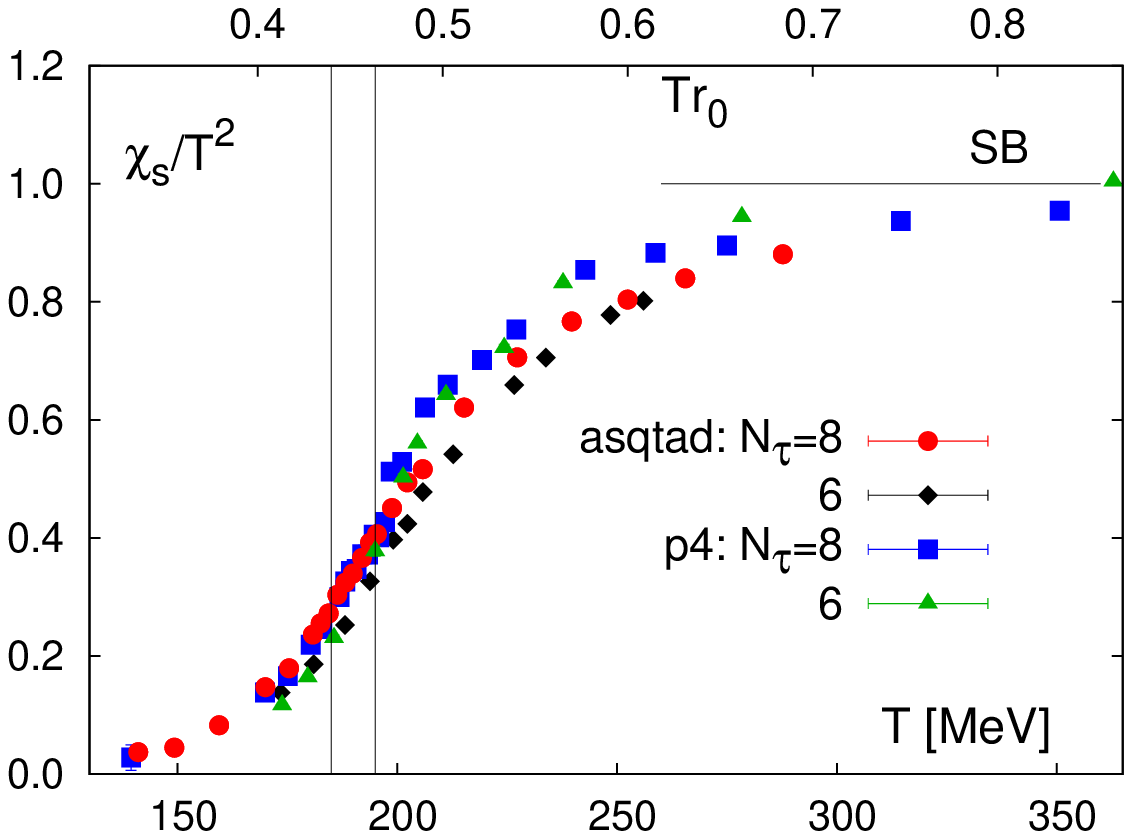}
\caption{On the left, the light quark number susceptibility.  On the right, the strange quark number
susceptibility.  For both quantities, the Stefan-Boltzmann limit is shown, as well as the crossover r
egion (185 MeV < T < 195 MeV).}
\label{fig:chi_q}
\end{figure}
\begin{figure}[hbt]
\includegraphics[width=0.47\textwidth]{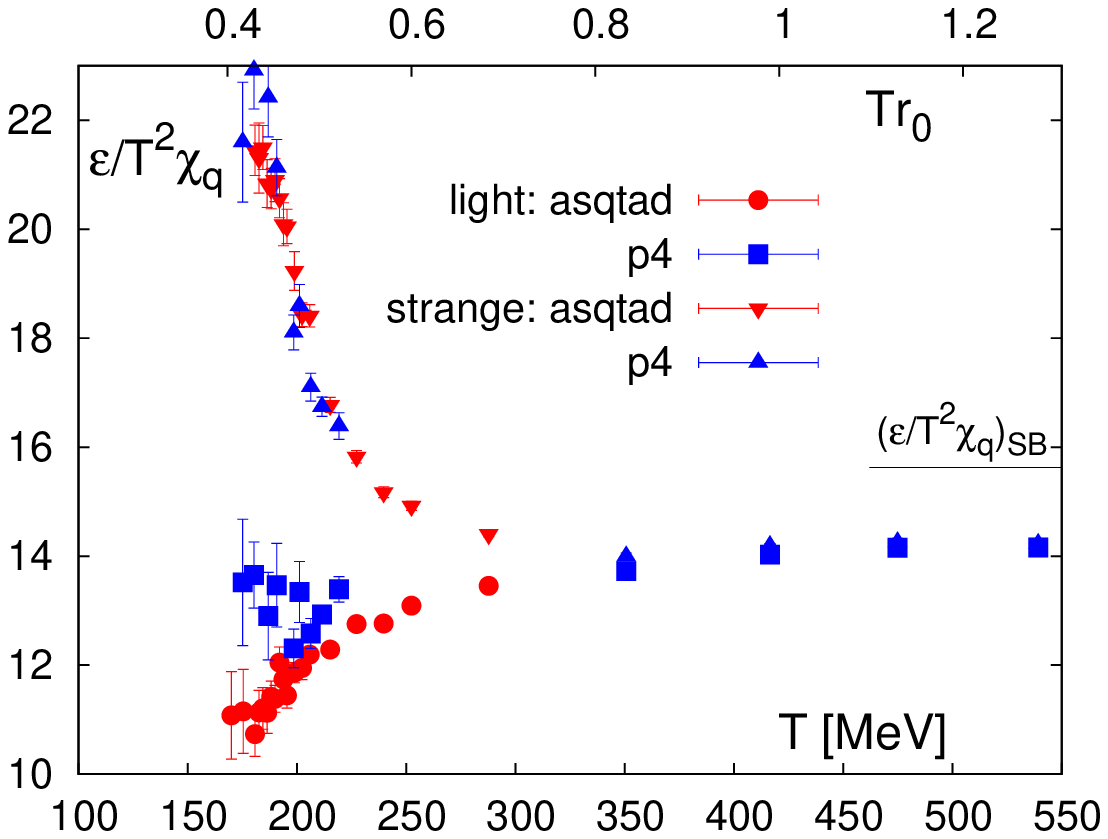}
\includegraphics[width=0.47\textwidth]{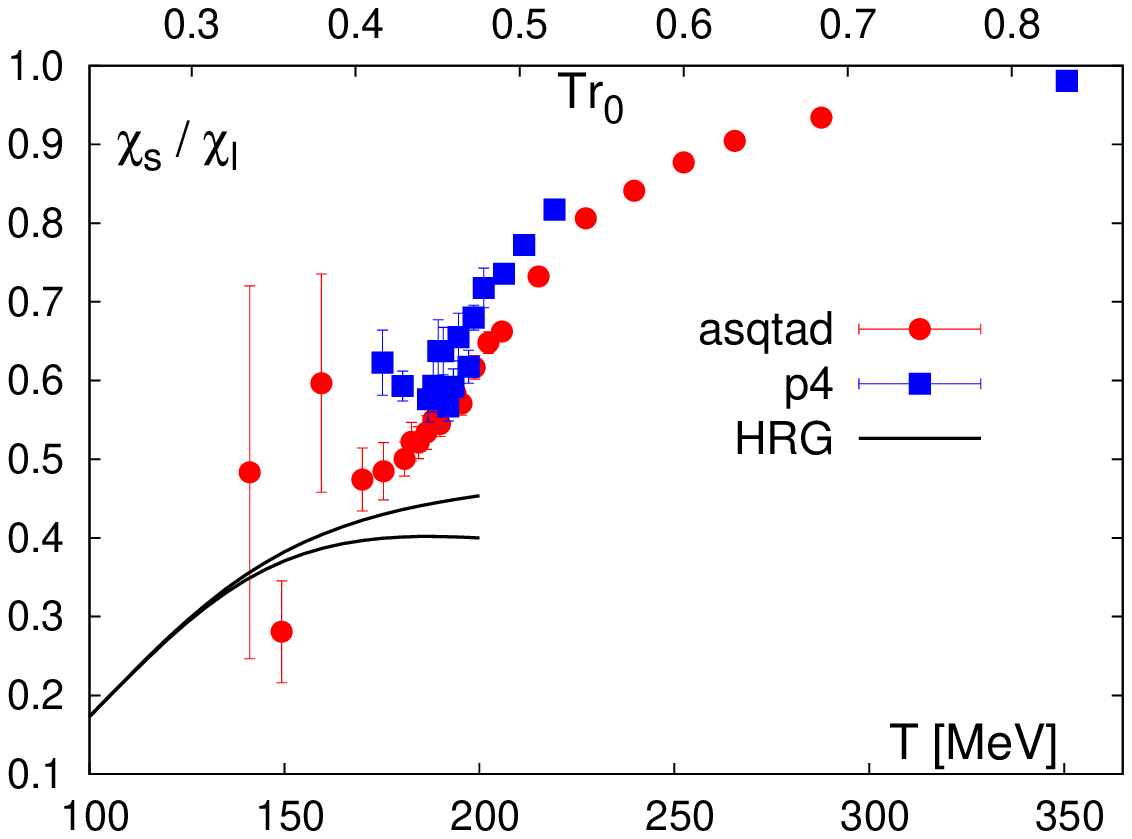}
\caption{On the left, the ratio of the energy density to the quark number susceptibilities.  On the 
right, the ratio of the strange to light quark number susceptibilities.  The solid
curves show this quantity in the hadron resonance gas model with $m_{max} = 1.5$ GeV 
(upper branch) and $m_{max} = 2.5$ GeV (lower branch).}
\label{fig:ratios}
\end{figure}
As it is suspected that QCD with physical quark masses undergoes a smooth crossover,
$T_c$ can not be sharply defined and may depend on the observable one chooses.  
In particular, it is unclear whether observables related to deconfinement
and those related to chiral symmetry restoration give similar values for $T_c$.  In this section,
we examine various quantities related to both deconfinement and chiral symmetry restoration.

One set of observables related to deconfinement are the light and strange quark number
susceptibilities:
\begin{equation}
\frac{\chi_q}{T^2} = \frac{1}{V T^3} \frac{ \partial^2 \ln Z}{\partial(\mu_q/T)^2}, ~ q = l,s.
\end{equation}
These quantities give information on the thermal fluctuations of the degrees of freedom that
carry net quark number.  Thus, this observable is sensitive to the
liberation of degrees of freedom that comes with deconfinement.  Figure \ref{fig:chi_q} shows
the light and strange quark number susceptibilities.  As we can see, $\chi_l/T^2$ rises more 
sharply than $\chi_s/T^2$, while $\chi_s/T^2$ seems to go to zero much more quickly.  This is
explained by the fact that the lightest hadrons that carry light and strange quark number are the
pions and kaons, respectively.  Thus, we would expect $\chi_l/T^2 \sim exp(-m_\pi/T)$, while
$\chi_s/T^2 \sim exp(-m_K/T)$. 

In Fig. \ref{fig:ratios},  we see that the light quark number susceptibility tracks the rise in
the energy density quite well as one moves through the crossover region.  On the other hand,
$\epsilon/T^2 \chi_s$ becomes singular in the low temperature regime.  This is because the
light quark susceptibility, but not the strange quark susceptibility, is directly sensitive to the 
singularity in the partition function in the chiral limit.  At high temperature, when the temperature
is sufficiently high that the quark masses are irrelevant, we see that $\epsilon/T^2 \chi_q$ agree.
The relative suppression of $\chi_s$ compared to $\chi_l$
can also be seen in the ratio $\chi_s/\chi_l$, also in Fig. \ref{fig:ratios}, where it is
consistent with the value obtained with the hadron resonance gas model.
\begin{figure}[hbt]
\includegraphics[width=0.47\textwidth]{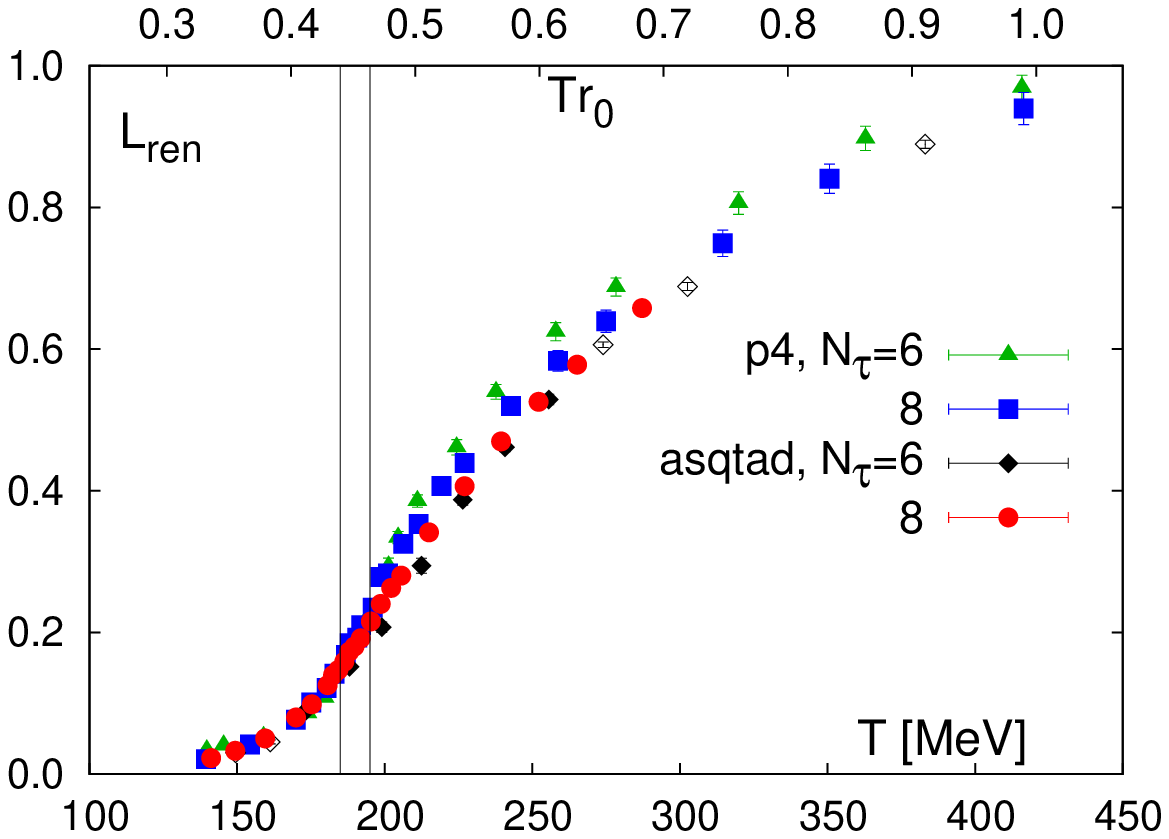}
\includegraphics[width=0.47\textwidth]{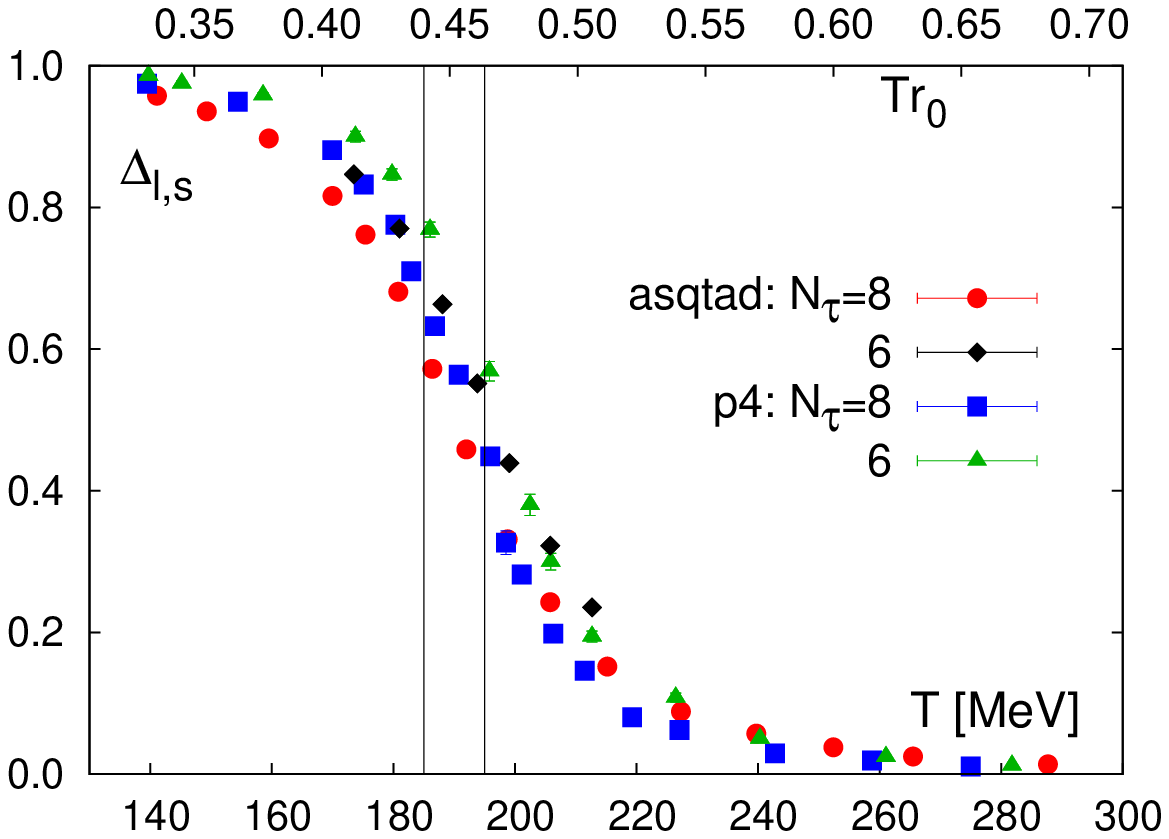}
\caption{On the left, the renormalized Polyakov for the two actions at $N_t = 6$ and 8.  On the right,
the subtracted chiral condensate, $\Delta_{l,s}$.}	
\label{fig:polyakov_pbp}
\end{figure}
In Fig. \ref{fig:polyakov_pbp}, we show the renormalized Polyakov loop and the subtracted
chiral condensate.  These quantities are sensitive to the chiral and deconfining transitions,
respectively.  The subtracted chiral condensate is defined as:
\begin{equation}
\Delta_{l,s}(T) = \frac{ \left< \bar{\psi} \psi \right>_{l,T} - \frac{m_l}{m_s} \left<\bar{\psi}\psi\right>_{s,T}}{\left<\bar{\psi}\psi\right>_{l,0} - \frac{m_l}{m_s} \left<\bar{\psi}\psi\right>_{s,0}}.
\end{equation}
The temperature band $185~\textrm{MeV} < T < 195~\textrm{MeV}$ is superimposed on both
plots to show that the chiral and deconfinement transitions occur in the same approximate 
temperature regime.

\section{Conclusions}
We have presented here a calculation of the bulk observables (energy density, pressure,
entropy density, speed of sound) of QCD matter at finite temperature and zero chemical
potential.  Our results indicate that cutoff errors are approximately 15\% in the
crossover region, and no more than 5\% for $T > 300~\textrm{MeV}.$  In addition, we have
calculated various observables (quark number susceptibility, renormalized Polyakov loop, and
chiral condensate), which seem to give consistent values of $T_c$ for the deconfinement and
chiral symmetry restoring transition at the lattice spacings that we employ.

\end{document}